\documentstyle[aps,preprint]{revtex}
\tightenlines

\begin{document}
\def\bi{\bigskip}
\def\be{\begin{equation}}
\def\en{\end{equation}}
\def\bq{\begin{eqnarray}}
\def\eq{\end{eqnarray}}
\def\noi{\noindent}
\def\bc{\begin{center}}
\def\ec{\end{center}}
\def\beit{\begin{itemize}}
\def\eit{\end{itemize}}
\def\ct{\centerline}

\title{Scalar Mesons and Chiral Dynamics}
\author{ Mauro Napsuciale$^1$\\ 
{\small \it Instituto de Fisica, Universidad de Guanajuato,\\
Lomas del Bosque 103, Facc. Lomas del Campestre \\
 37150, Leon, Guanajuato, Mexico.}}

\maketitle


\begin{abstract}
We discuss scalar mesons properties on the light of chiral dynamics. Considering 
them as the chiral partners of pseudo-scalar mesons we propose an explanation to their 
unusual 
properties based on non-trivial vacuum effects coming from the interplay between 
spontaneous breaking of chiral symmetry and the violation of $U_A(1)$ symmetry by 
instantons. Including vector mesons as external sources we work out predictions 
for radiative decays of vector mesons and compare some of them with recent 
experimental results from high luminosity $\Phi$ factories.
\end{abstract}

\maketitle

\section{Introduction}
The understanding of scalar excitations is a fundamental problem which we 
encounter in many branches of physics ranging from the dilaton in theories for 
gravity, the higgs in the 
electroweak theory, to the pairing of fermions in condensed matter. The reason 
is quite simple: scalar excitations have the same quantum numbers as the 
vacuum and their properties are strongly influenced by it. Understanding 
scalar excitations is at the same time, in some way,  an understanding of the 
vacuum properties of the corresponding theory which is a particularly 
difficult task in the case it has an strongly coupled regime. 

In this talk I will summarize our work on scalar mesons in low energy 
QCD. The well-established lowest lying scalar mesons are the isovector $a_0(980)$ 
and the isoscalar $f_0(980)$. These mesons are nearly degenerate in mass which
 suggests they are the scalar analogous to the $\rho (770)$ and $\omega (780)$ 
which would imply a $\frac{1}{\sqrt{2}}(\bar u u-\bar dd )$ and  
$\frac{1}{\sqrt{2}}(\bar u u+\bar dd )$  for the  $a_0(980)$ and $f_0(980)$ 
respectively. The first problem with this identification is the 
experimental fact that the $f_0(980)$  strongly couples to the $\bar KK $ 
system. The second problem is the small coupling to two photons these mesons have, 
a problem which we will review in detail below. Before this let us mention that 
over the past few years compelling evidence has accumulated for the existence of 
a broad isoscalar structure ($\sigma$) in the very low energy region \cite{tornq} 
which is strongly coupled to two pions and the existence of a isospinor 
scalar in the $800-900$ MeV has been claimed by many authors \cite{kappa}, although 
its existence is still under debate \cite{penn}.

\subsection{Scalar (and pseudoscalar) mesons as $\bar QQ$ states: testing the 
structure of hadrons with photons.}

Electromagnetic decays of hadrons give valuable information on their structure 
for the simple reason that photons couple to charged objects, in particular 
to the partons constituting a hadron. In this sense the two photon decay of 
scalars constitute direct evidence for their quark structure. Quark model 
calculations for the $a_0(980)\to\gamma\gamma$ and $f_0(980)\to\gamma\gamma$ 
decays were summarized in \cite{barnes} and a molecule structure was explored 
in this work which seemed to be favored by the measured branching ratios. Here 
we will work with the following assumptions: i) meson are composed of  $\bar QQ$ 
where $Q $ 
denotes a {\bf constituent} (colored) quark. ii)They are non-relativistic 
systems and we work in the zero-binding approximation ($M\approx 2m_Q$). 
Under these assumptions we can 
use either NRQCD-like calculations (singlet channels only) or we can apply 
the quarkonium techniques developed in \cite{germans}. The calculations lead to 
\[
\begin{array}{lcl} 
\Gamma(^1S_0\to\gamma\gamma)=\frac{12\alpha^2}{M^2_P}~|R(0)|^2 ~e^2_Q \\
\Gamma(^3P_J\to\gamma\gamma)=\frac{N_J\alpha^2}{M^4_J}~|R^\prime (0)|^2 ~
e^2_Q 
\end{array} \nonumber
\]

\noi where $N_0=432,~ N_2=576/5$ and $N_1=0$ as required 
by charge conjugation. Here, $e_Q$ denotes the charge of the constituent 
quark (in units of $e$) and $R(0)$, $R^\prime (0)$ denote the quarkonium 
wave funcion and its derivative evaluated at the origin respectively.
For the physical process {\it meson} $\to\gamma\gamma$ it is necessary to 
consider the isospin structure which amounts to replace $e^2_Q\to F(I)$, where 
$F(1)=(e^2_u-e^2_d)/\sqrt{2}$ for a 
$(\bar uu -\bar dd)/\sqrt{2}$ meson and   
$F(0)=(e^2_u+e^2_d)/\sqrt{2}$ for a 
$(\bar uu +\bar dd)/\sqrt{2}$ meson.
The only unknown parameters here are the wave functions at the origin which 
do not allow us to predict the individual widths. However, this factor 
cancels out in the ratios of widths which allow us to compare with the 
measured ratios for mesons with the same orbital angular momentum. 
We show in a table the results for the ratios of different combinations of 
the $^3P_J \to\gamma\gamma$ decay widths 
$R^{M_1M_2}_{th}\equiv \frac{\Gamma(M_1\to \gamma\gamma )}
{\Gamma(M_2\to \gamma\gamma )}=(\frac{m_{M_2}}{m_{M_1}})^4 \frac{F^2(I_1)}
{F^2(I_2)} $. In this table a $(\bar uu+\bar dd )/\sqrt{2}$ 
structure has been assumed for the $f_0$ and $f_2$ mesons and a 
$(\bar uu-\bar dd )/\sqrt{2}$ structure for the $a_0$ and $a_2$ 
mesons.

\bc
\begin{tabular}{|c|c|c|} \hline
$M_1M_2$          & $R^{M1M2}_{th}$& $R^{M1M2}_{exp}$   \\ \hline
$a_2(1320)f_2(1270)  $  & $ 0.31 $ & $ 0.36 \pm 0.05 $  \\ \hline
   $a_0(980)f_0(980) $  & $ 0.35 $ & $ 1.62 \pm 0.96 $  \\ \hline
   $f_0(980)f_2(1270)$  & $ 0.76 $ & $ 0.14 \pm 0.04 $  \\ \hline
   $a_0(980)a_2(980) $  & $ 0.85 $ & $ 0.24 \pm 0.09 $  \\ \hline
\end{tabular}
\ec
The conclusion we extract from this table is that two photon decays of 
tensor mesons are consistent with the assumed composition whereas
{\bf $a_0(980)\to \gamma\gamma $ and $ f_0(980) \to \gamma\gamma$
are not consistent with a 
$(\bar uu \pm \bar dd )/\sqrt{2}$ structure for these mesons}. Another possibility 
in the case of the  $f_0(980)$ is a $\bar ss$ structure. For this case we 
obtain $R^{f_0f_2}_{th} =0.06$ vs $R^{f_0f_2}_{exp}= 0.14\pm 0.04$. Thus 
 $ f_0(980)\to\gamma\gamma  $ is neither  well 
described in terms of quarkonium calculations assuming a pure $\bar ss$ 
structure. It still remain the possibility that the $f_0(980)$ be a more 
complicated object similar to the $\eta$ or $\eta^\prime$ i.e. 
$f_0= \sin \phi_{\rm s}S_{\rm ns}+\cos\phi_{\rm s} S_{\rm s} $
where $S_{\rm ns}= (\bar uu+\bar dd)/\sqrt{2}$ and $S_{\rm s}= 
\bar ss$. In this case 
$ R^{f_0f_2}_{th}= \frac{4}{15}(\frac{M_{f_2}}{M_{f_0}})^4 
(\sin \phi_{\rm s}+\frac{\sqrt{2}}{5}\cos\phi_{\rm s}) 
$ and agreement with the experimental ratio requires 
$\phi_{\rm s} \approx 9^\circ $. 
 
What about pseudoscalars?. In this case the quarkonium state has the 
$J^{PC}=0^{-+}$ quantum numbers ($^1S_0$ state in spectroscopic notation).
The predicted ratio of widths in the case $\eta-\pi$, written in terms of 
the mixing angle in the singlet-octet basis $\theta_P=\phi_P -54.7^\circ$  is 
$R^{\eta\pi}_{th}= \frac{m^3_\eta}{m^3_\pi}(\frac{m_\pi}{m_\eta})^5 
\frac{1}{\sqrt{3}}\lbrack\cos\theta_{\rm p}-2\sqrt{2}\sin\theta_{\rm p})
\rbrack ^2 $. In this case, the results for the ratio of widths differ from 
those arising in  naive quark model (or SU(3))  (see e.g. \cite{RPP} page 117) .
Indeed, there is a  dynamical factor 
$(\frac{m_\pi}{m_\eta})^5 \approx 10^{-3}$ for this ratio which considerable 
reduce the ratio with respect to naive quark model considerations when 
physical masses are used in the numerics. Reproducing experimental data 
require to introduce a ``reference mass''($M_R$) \cite{isgur} whose origin 
is clear in our framework, $M_R\approx 2m_Q$ which represent the 
``constituent mass'' for the quarkonium. The inability of quarkonium 
calculations to describe two photon decays of pseudoscalars without any 
additional assumption has to do with the mechanisms which deviate the 
values of the masses of pseudoscalars from their ``constituent mass'' values. 
It has been clear since the early days of QCD that the spectrum 
of pseudoscalars is strongly influenced by non-trivial effects due to the 
properties of the QCD vacuum. On the other hand scalars have the same quantum numbers
as the vacuum and the immediate question is whether or not the scalar spectrum 
and decay properties are modified, from what we naively would expect, by 
effects due to the properties of the QCD vacuum. In the next section we 
will review our work on the effects of the vacuum in the structuring of 
the spectrum and decay properties of scalar mesons.

\subsection{Chiral symmetry, vacuum effects and scalar meson properties.}

We explored this topic in the framework of a Linear Sigma Model (LSM) which 
incorporates the most important effects in the low energy domain of QCD, 
namely, the spontaneous breaking of $\chi$ral symmetry (SB$\chi$S) and the 
breakdown of the singlet axial symmetry. For the latter 
we use the bosonized version of the instanton induced six-quark interaction 
discovered by 't Hooft. The pseudoscalar and
scalar matrix fields $P$ and $\sigma $ are written in terms of a specific
basis spanned by seven of the standard Gell-Mann matrices, namely $\lambda_i
~(i=1,\ldots , 7)$, {\em and} by two unconventional matrices 
$\lambda_{\rm{ns}}$=diag(1,1,0), and $\lambda_{\rm{s}}$ = $\sqrt{2}$
diag(0,0,1), respectively. We use the convention  
$P\equiv \frac{1}{\sqrt{2}} \lambda_i P_i $ with $i=ns,s,1,\ldots,7$ and 
similarly for the scalar field. The  scalar and a pseudoscalar
mesonic nonet enter in the chiral combination $M=\sigma + i P$. The 
lagrangian  is
\begin{equation}
{\mathcal L}= {\mathcal L}_{sym}  + {\mathcal L}_{U_A(1)} + 
{\mathcal L}_{SB}.
\end{equation}

The $ \lbrack U(3)_L\otimes U(3)_R\rbrack $ symmetric part is given by 

\be
 {\mathcal L}_{sym}=\frac{1}{2}{\rm tr}\left[(\partial_\mu M)
 (\partial^\mu M^{\dagger}) \right]  -\frac{\mu^2}{2} 
 X (\sigma ,P)
   -\frac{\lambda}{4} Y(\sigma ,P)
 -\frac{\lambda^\prime}{4} X^2 (\sigma , P)  , 
 \label{symmlag}
\en
where $X(\sigma,P)\equiv{\rm tr}\left[ M M^{\dagger}\right]~~  
Y(\sigma,P)\equiv{\rm tr}\left[ (M M^{\dagger})^2\right ]$. The $U_A(1)$ 
symmetry breaking lagrangian is given by $ {\mathcal L}_{U_A(1)} = 
- \beta \lbrace  \mbox{det}(M)+ \mbox{det} (M^{\dagger }) \rbrace $. 
Finally we have the explicit symmetry breaking term $ {\mathcal L}_{SB} =  
{\rm tr} \left[ c\sigma \right]= 
{\rm tr} \left[ \frac{b_0}{\sqrt{2}}{\cal M}_q(M+M^{\dagger}) \right]$ where 
$c\equiv \frac{1}{\sqrt{2}} \lambda_i c_i$,is related to the
quark mass matrix ${\cal M}_q$ by $c=\sqrt{2}b_0{\cal M}_q$ and has  
$\frac{c_{\rm{ns}}}{\sqrt{2}}= \sqrt{2}\hat m b_0$ and $c_{\rm{s}}=
\sqrt{2} m_s 
b_0$ as the only non-vanishing entries. Here, $b_0$ is an unknown parameter 
with dimensions of squared mass. We work in the exact isospin limit,
 $\hat m =m_u=m_d$. The linear $\sigma$ term induces $\sigma$-vacuum 
transitions which rearrange the vacuum. Shifting to physical fields we 
obtain the following masses \cite{mauro,mariana,simon,tornq2,thooft} :
\[
 m^2_\pi =\xi \underline{+2\beta b}+\lambda a^2, \qquad  ~~
 m^2_{a_0} = \xi \underline{ -2\beta b} +3\lambda a^2  
\]
\[ 
m^2_K =\xi \underline{ +2\beta a}+\lambda (a^2\mbox{$-$}ab\mbox{+}b^2), ~~~ 
 m^2_\kappa =\xi \underline{ -2\beta a}+\lambda (a^2\mbox{$+$}ab\mbox{+}b^2)
\]
\be
  m^2_{\eta_{\rm{ns}}} =\xi \underline{ -2\beta b} + \lambda a^2 \;,  \qquad 
  m^2_{S_{\rm {ns}}} =\xi \underline{ +2\beta b} + 3\lambda a^2 + 
4\lambda^\prime a^2, \label{masses}
\en
\[ 
 m^2_{\eta_{\rm {s}}} =\xi + \lambda b^2 \; ,   \qquad  
 m^2_{S_{\rm{s}}} =\xi + 3\lambda b^2 + 2\lambda^\prime b^2 
 \]
\[
 m^2_{\eta_{\rm {s-ns}}} =\underline{ -2\sqrt{2}\beta a} \; ,  \qquad  
 m^2_{S_{\rm{s-ns}}} = \underline{ 2\sqrt{2}}(\underline{\beta} 
+\lambda^\prime b)\underline{a} \;.
\]
\noi where $a=\langle \sigma_{\rm{ns}}\rangle/\sqrt{2}$, 
$b=\langle \sigma_{\rm{s}}\rangle$ and $m^2_{\eta_{\rm {s-ns}}}$, $ 
m^2_{S_{\rm{s-ns}}}$ denote the OZI 
rule violating terms mixing strange and non-strange isoscalar quarkonia. 
Notice that the $U_A(1)$ symmetry breaking couples to   
the spontaneous breakdown of chiral symmetry (underlined terms in 
Eq.(\ref{masses})) contributing to the masses of 
all fields (except strange fields) and this effect has the opposite sign in 
the scalar sector with respect  to the pseudoscalar sector. 
In particular it is responsible for the mixing of flavor fields. 
We claim that this is the striking effect which explains the unusual 
properties of the lowest lying scalar mesons. In \cite{mauro,mariana} 
the corresponding coupling has been fixed to 
$\beta= -1.551\pm 0.072 $ GeV using information on the pseudoscalar 
spectrum as input and the members of the scalar nonet were identified 
as $\sigma(\approx 450),~ f_0(980),~a_0(980),$ and $\kappa(\approx 900)$. The 
$U_A(1)$-SB$\chi$S effect has the following consequences: 
It pushes the pions and kaons {\bf down} making them light 
(an effect driven by the quark masses since elimination of linear terms after 
SB$\chi$S requires $f_{\pi} m^2_\pi= 2 \hat m b_0 ;~~~f_{K} m_K^2 = 
(\hat m +m_{\rm s})b_0$ ) and the $\eta_{\rm ns}$ {\bf up} making it 
heavy (an effect only partly driven by quark masses).ii) It pushes the $a_0$ 
and $\kappa$ mesons {\bf up} making them heavy and 
$\sigma_{\rm ns}$ {\bf down} making it light. As a consequence this effect 
simultaneously explains at the qualitative level :i) Why the pion and kaon 
are light, ii) The mixing of pseudoscalar and scalar mesons.
iii) Why the $\eta_{\rm ns}$ is so heavy, 
iv) Why the sigma meson is so light, vi) 
The accidental degeneracy of the $a_0(980)$ and the $f_0(980)$ 
(and perhaps also of the $\kappa$ meson) 
vi) The strong coupling of the $f_0(980)$ to $\bar KK$.
The immediate question at this point is what the model predicts for the 
coupling of $f_0(980)$ and $a_0(980)$ to two photons.

\subsection{ Consistent description of $a_0(980)\to \gamma\gamma $ 
and $f_0(980)\to \gamma\gamma$ decays}

The $a_0(980)\to \gamma\gamma$ and  $f_0(980)\to \gamma\gamma$ decays have 
been calculated in the present framework \cite{luna}. These decays are  
induced by loops of charged mesons ($M$). The calculation of the 
corresponding diagrams give {\it finite} contributions. The analytical 
results depend on the $SMM$ couplings which are related to meson masses 
by chiral symmetry. The amplitude for $S\to \gamma\gamma$ decay is
\be
{\mathcal M} (S\to\gamma (\varepsilon, k) \gamma (q,\eta ))= 
\frac{i\alpha}{\pi f_{ K}}{\mathcal A}^{ S} ( q.k ~g^{\mu\nu}- k^{\mu}q^{\nu}~)
\eta_{\mu}~\varepsilon_{\nu}.
\en
\noi and the corresponding width is $
\Gamma(S \to\gamma\gamma) = \frac{\alpha^2}{64\pi^3} \frac{m^3_{S}}
{f^2_{ K}}|{\cal A}^{ S}|^2$, whereas the measured widths are \cite{RPP} 
$\Gamma(f_0\to\gamma\gamma)_{exp}=  0.39^{+10}_{-13}~ \mbox{keV},  ~~~ 
\Gamma (a_0\to \gamma\gamma)_{exp}= 0.24^{+0.08}_{-0.07}~ \mbox{keV} $ 
(we assume 
$ BR(a_0(980)\to \pi^0\eta)\cong 1$). From these values we 
extract $|{\cal A}^{a_0}_{exp}|= 0.34 \pm 0.05$ and  
$|{\mathcal A}^{f_0}_{exp}| = 0.44\pm 0.07 $ . 

In the model, the  $a_0\to\gamma\gamma$ 
decay is induced by loops of kaons and kappas. Kaon loops yield 
${\mathcal A}^{a_0}_{_K} = 0.42 $. Kappa loops interfere destructively  
with kaon 
loops and results depend on the kappa mass. Using  e.g. $m_\kappa=900$~MeV
we obtain
${\cal A}^{a_0}_{_{LSM}}=0.36 $. The experimental result is consistent with 
$m_\kappa\in [800, 935] $~MeV. 

In the case of the $f_0(980)\to \gamma\gamma$ decay there are contributions 
from loops of pions kaons and kappas. The corresponding amplitudes as 
calculated in the model yield 
${\cal A}^{f_0}_{_M} =f_{_K} \left( \frac{g_{_{f_{0}MM}} }{m^2_{f_0}} 
\right) N_{_M}$ with the loop factors $N_\pi =(- 1.10+0.48 i)$,  
$N_{_K} = 1.06$ and 
the value of $N_\kappa$ depend again on the kappa mass. For  
$m_\kappa=900$~MeV we obtain $N_\kappa =0.12$. The dominant contribution 
again comes from kaon loops since the $f_0\pi\pi$ coupling is proportional 
to the $\sin\phi_{\rm s}$ factor 
which is small, and the $f_0\kappa\kappa $ coupling is proportional to 
$m^2_f-m^2_\kappa$ which is also small relative to the typical hadron 
energy scale of 1 GeV. Kaon loops contributions alone 
(using $\phi_s=-9^\circ$) yield ${\cal A}^{f_0}_{_K}= 0.46 $ whereas 
including all contributions (using the E791 
value $m_\kappa= 797$~MeV  \cite{E791-2}) yield  
$|{\cal A}^{f_0}_{_{LSM}}| =0.42$ to be compared with 
$|{\cal A}^{f_0}_{exp}|= 0.44\pm 0.06 $. Reversing the argument: 
fixing the value of $m_{\kappa}$ to the central value of E791, the 
experimental uncertainties allow for $\phi_s \in [-4^\circ,-25^\circ ]$. 
In the whole we conclude that experimental data on  $a_0\to \gamma\gamma$ 
and $f_0\to\gamma\gamma$ are well described by meson loops. 

\subsection{ Complementarity of LSM and $\chi$PT in the scalar channel: 
$ V^0 \to P^0{P^0}^\prime\gamma $ decays.}

From the experimental point of view these decays are a clean place to study 
$P^0-{P^0}^\prime$ systems since neutral particles are involved in the final 
state, hence there exist no final state radiation. Branching ratios and energy 
spectrum were recently measured for some of 
these decays by the SND and CMD collaborations at  Novosibirsk and 
improved data 
can be expected from DA$\Phi$NE. On the theoretical side contributions from 
intermediate vector mesons $V^0\to {V^0}^\prime P^0\to P^0{P^0}^\prime 
\gamma$ were 
estimated in \cite{bramon1} using VMD and the corresponding results 
do not describe 
the experimental results for the BR.'s. The possibility of 
enhancement of these BR's due to re-scattering effects was explored in 
\cite{bramon2} 
using $\chi$PT with vector mesons as external fields. At leading order 
(${\cal O}(p^4)$) these contributions are  {\bf finite} and no 
ambiguities due to 
counter-terms exist. VMD and chiral 
loops contributions are collected in a table below which shows that these 
contributions do not account for the measured BR's either. Since the 
$P^0-{P^0}^\prime$ system can be in a $J^P=0^+$ state we should expect 
resonant effects due to scalars manifest in these decays.

\begin{tabular}{|c|c|c|c|} \hline
Decay & BR$_{\rm exp}\times 10^5$ & (VC)$\times$ 10$^5$ &
($\chi$PT) $\times$ 10$^5$ \\ \hline
$\phi \to\pi^0\pi^0\gamma$ & 11.58 $\pm$ 0.93 $\pm$ 0.52 & 1.2 & 5.05  \\ 
\hline
$\phi\to\pi^0 \eta\gamma$ & 9.0 $\pm$ 2.4 $\pm$ 1.0 & 0.54 & 2.95  \\ \hline
$\phi\to K^0\bar K^0\gamma$ & & 2.7 $\times$ 10$^{-7}$ & 1.0 
$\times$ 10$^{-3}$ \\ \hline
$\rho\to\pi^0\pi^0 \gamma $ & 4.2$^{+2.9}_{-2.0}\pm$ 1.0 & 1.1 & 0.97 
\\ \hline 
$\rho\to \pi^0 \eta\gamma$ & & 4 $\times$ 10$^{-5}$ 
&  4 $\times$ 10$^{-6}$ \\ \hline
$\omega\to \pi^0\pi^0 \gamma$ & 7.2 $\pm$ 2.5 
& 2.8 & 9 $\times$ 10$^{-2}$ \\ \hline
$\omega\to\pi^0 \eta\gamma$ & & 1.6 $\times$ 10$^{-2}$ & 1.6 $\times$ 
10$^{-4}$ 
\\ \hline
\end{tabular} 

Vector fields were introduced as external field in the model in 
\cite{nos1,nos2,nos3} and contributions of intermediate scalar mesons to these 
processes were calculated for the most interesting processes, namely  
$\phi\to\pi^0\pi^0\gamma$, $\phi\to\pi^0\eta\gamma$ and 
$\rho,\omega\to\pi^0\pi^0\gamma$. We refer to \cite{nos1,nos2,nos3} for 
details. Here we just summarize the main results. From the theoretical point 
of view is worth to remark that the obtained amplitudes to these processes 
reduce to the ${\cal O} (p^4)$ $\chi$PT amplitudes obtained in 
\cite{bramon2} in the case of heavy scalars. In this sense, 
LSM yields results which are complementary to $\chi$PT. 
In addition to catch the physics of chiral loops the LSM amplitude   
is also able to reproduce the effects of the scalar poles at higher  
$P^0{P^0}^\prime$ invariant mass values. In the case of 
$\phi\to P^0 {P^0}^\prime \gamma $ pion loops are suppressed by G-parity, 
hence kaon loops give the most important contribution. 
The decay $\phi\to \pi^0\pi^0\gamma $ is highly sensitive to the scalar 
mixing angle and it 
is not sensitive to the $\sigma$ mass whenever it be light since 
$g_{\sigma KK}\sim m^2_\sigma -m^2_K $. Disastrous results are obtained for 
$m_\sigma > 600$~MeV. The energy spectrum for this process is nicely 
reproduced by the calculations in the model \cite{nos1}. The calculated 
branching ratios, including intermediate vector meson contributions is 
$BR(\phi\to\pi^0\pi^0\gamma)_{\mbox{TH}}= 1.08\times 10^{-4}$ which is to be 
compared with $BR(\phi\to\pi^0\pi^0\gamma)_{\mbox{EXP}}= 
1.08\pm0.17\pm0.09 $ measured by the CMD2 coll.\cite{CMD2} and 
 $BR(\phi\to\pi^0\pi^0\gamma)_{\mbox{EXP}}=1.14 \pm 0.10\pm 0.12$ obtained by the SND 
Coll. \cite{SND}.

In the case of $\phi\to\pi^0\eta\gamma$, the only intermediate scalar 
is the $a_0(980)$\cite{nos2}. The corresponding spectrum 
is also reproduced in the model. The branching ratio calculated in the 
model is 
$B(\phi\rightarrow\pi^0\eta\gamma)_{\rm LSM}= (0.75$--$0.95)\times 10^{-4}$, 
to be compared with the values reported by the CMD2 Collaboration 
$B(\phi\rightarrow\pi^0\eta\gamma)_{\rm CMD2}=(0.90\pm 0.24\pm 0.10)
\times 10^{-4}$ \cite{CMD2} and the SND result 
$B(\phi\rightarrow\pi^0\eta\gamma)_{\rm SND}=
(0.88\pm 0.14\pm 0.09)\times 10^{-4}$ \cite{SND1}.
Improved data near the $a_0$ pole will be very important in the 
understanding of the $a_0(980)$ meson.

In the case of $\rho\to\pi^0\pi^0\gamma$ exchange of vector mesons account for 
$\approx 25\% $ of the measured BR, $\chi$ral loops (dominated by pions) 
account for another $\approx 20\% $ and the calculated BR taking into 
account both contributions \cite{bramon2} is within two standard deviations 
from the experimental results recently reported by the SND Coll. 
\cite{SND1} and quoted in table above. An analysis of the energy spectrum 
as a function of the mass and width of the $\sigma$ meson shows that this 
process is sensitive enough to these quantities and a measurement of the 
spectrum can be used to extract the corresponding values. Integrating the 
$\pi^0\pi^0$ invariant mass spectrum and using  the central values 
obtained by the E791 Collaboration  $m_\sigma=478^{+24}_{-23}\pm 17$ MeV and 
$\Gamma_{\sigma}=324^{+42}_{-40}\pm 21$ MeV we obtain $
BR(\rho\rightarrow\pi^0\pi^0\gamma)_{\mbox{\rm LSM}}=
1.5\times 10^{-5}$ .For $m_\sigma=478$ MeV and a narrower width 
$\Gamma_\sigma=263$ MeV (as predicted by the LSM, see also \cite{BEPC}) we 
obtain  $BR(\rho\rightarrow\pi^0\pi^0\gamma)_{\mbox{\rm LSM}}=
2.1\times 10^{-5}$. 

Finally, in the case of  $\omega\to \pi^0\pi^0\gamma$ pion loops are also 
suppressed by G-parity and kaon loops should give the most important 
contribution. However, the intermediate scalars are 
$f_0(980)$ and $\sigma$ and the $f_0(980)$ has 
a large mass compared to the energy region of interest whereas 
$\sigma$ contributions are highly suppressed since  
$g_{\sigma KK}\sim m^2_\sigma -m^2_K $. Hence, whenever 
the mass of the sigma meson be small, scalar effects are negligible in 
this process, although interference with the intermediate vector meson 
contributions seem to close the gap between the experimental results  
$BR(\omega\to\pi^0\pi^0\gamma)_{\rm \mbox{EXP}}= (7.2\pm 2.5 )\times 10^{-5}$ 
\cite{RPP} and the VM contributions when we take care of including 
properly the $\omega-\rho$ mixing \cite{nos3} which yield 
$BR(\omega\to\pi^0\pi^0\gamma)_{\mbox{TH}}= 
4.5\pm 1.1 \times 10^{-5}$ \cite{nos3} 
(see also \cite{singer}).

\subsection{Conclusions.}
Summarizing, calculations for  $S\to\gamma\gamma$ ($S=a_0(980),f_0(980)$) 
considering $S$ as a NR-quarkonium state yield results which are not 
consistent with experimental data. The same is true in the case of 
pseudoscalars. In this case the disagreement can be traced back to the strong 
distortion of the spectrum from naive quarkonium expectations due to QCD 
vacuum effects. Since scalars have the same quantum numbers as the vacuum, 
highly non-trivial effects are expected in this sector 
due to vacuum properties. We study such effects in the framework of a 
phenomenological $U(3)\times U(3)$ chiral lagrangian which incorporates 
spontaneous $\chi$S and $U_A(1)$ symmetry breaking. In this framework 
there is an important effect: the coupling of the $U_A(1)$ violating 
interaction to the spontaneous breaking of chiral symmetry which generates 
mass terms ($U_A(1)$-SB$\chi$S effect).
This effect simultaneously explains: the smallness of the masses of the 
pions and kaons (an effect driven by quark masses), the OZI-rule violating 
mixing of flavor fields, the accidental degeneracy of the 
$a_0(980)$ and $f_0(980)$ , the lightness of 
the sigma meson and the controversial $a_0,f_0\to \gamma\gamma$ 
decays. Intermediate scalar contributions to $V^0\to P^0{P^0}^\prime\gamma$ 
are also calculated.  The energy spectrum of $\phi\to\pi^0\pi^0\gamma $ is 
nicely described giving direct evidence for the assignment of the 
$f_0(980)$ and indirect evidence for a light $\sigma$. Calculations for 
total and partial BR's are in good agreement 
with measurements from SND and CMD collaborations. 
Energy spectrum in $\phi\to\pi^0\eta\gamma $ is well described by the 
chiral loops + scalar resonant effects. Improvement in the measurements 
near the $a_0$ pole are encouraged. We also calculate the 
$\rho\to\pi^0\pi^0\gamma $ decay.
The corresponding BR is sensitive to the parameters of 
the $\sigma$. The branching ratio as measured by the SND Coll. 
is consistent with the results of the model when the values for the mass 
and width of this scalar as measured by the E791 Collaboration are used. 
Measurements of the energy spectrum are encouraged. 
Finally, $\omega \to \pi^0\pi^0\gamma $ is not sensitive to 
scalar contributions.

\subsection{Acknowledgements}

I wish to thank J. L Lucio, S. Rodriguez, A. Bramon, R. Escribano, 
M. Kirchbach and A. Wirzba for an enjoyable collaboration. 
This work was supported by CONCYTEG, Mexico under contract 
00-16-CONCYTEG-CONACYT-075.

\end{document}